\documentstyle[preprint,aps]{revtex}
\draft
\begin{document}

\tightenlines

\title{Semiclassical description of resonant tunneling}
\address{E.B. Bogomolny and D.C. Rouben\\
Division de Physique Th\'eorique\cite{00},
Institut de Physique Nucl\'eaire\\
91406 Orsay Cedex, France}

\maketitle

\begin{abstract}
We derive a semiclassical formula for the tunneling current of electrons
trapped in a potential well which can tunnel into and across a wide quantum
well.  The  calculations idealize an experimental situation where a strong 
magnetic field tilted with respect to an electric field is used. 
The resulting semiclassical expression is written as the sum over 
special periodic orbits which hit both walls of the quantum well and are 
perpendicular to the first wall.
\end{abstract}



The connection between quantum properties and the underlying classical
mechanics has attracted wide attention in the last years (see e.g. \cite{4} 
and references therein). The central result of this investigation, the 
Gutzwiller trace formula, relates long-range fluctuations in the density of
quantized levels with classical periodic orbits. Each periodic orbit of the
classical dynamical system with period $T_p$ generates regular maxima in the
level density separated by an energy $\Delta E_p=\hbar/T_p$. 

The influence of classical motion on quantum mechanical systems has 
also been investigated experimentally \cite{6}-\cite{jalab}. In all cases the 
observed oscillations were attributed to fluctuations  due to special periodic 
orbits.

In this paper we discuss a particular system to which much experimental 
work has been devoted in recent years \cite{8a} -\cite{9b}. The double well 
potential consisting of 
GaAs/(AlGa)As resonant tunneling diode containing a wide quantum well (QW) 
has been used to explore a relationship between classical and quantum
electron dynamics. In the presence of a strong uniform magnetic
field ($B$) tilted with respect to the $z$-axis by an angle $\theta$ and a 
uniform electric field ($\epsilon$) parallel to the $z$-axis, the system 
exhibits chaotic
motion for certain initial conditions in phase space \cite{8a} -\cite{ms} and
the experiments and numerical simulations reveal oscillations in the tunneling 
current associated with certain classical periodic orbits \cite{8a} -\cite{ms}.
Even the scarring of wave functions for this problem has been found \cite{9b}.

The main purpose of this paper is to obtain a quantitative description 
of the link between the tunneling current and periodic orbits. We shall show 
that  in the strict semiclassical approximation the main contribution will be 
from special periodic orbits which (i) hit the two walls and (ii) are 
perpendicular to the first barrier. Unlike the Gutzwiller trace formula the 
result is not a canonical invariant.

The Hamiltonian (in atomic units) for motion of an electron inside the system, 
is effectively 2-dimensional and can be written as:
\begin{equation}
H(y,z)=\frac{\vec{p}\;^2}{2m}+\frac{B^2}{2m}(y\cos\theta-z\sin\theta)^2
-{\varepsilon}z + U(z),
\label{ham}
\end{equation}
where $U(z)$ is a step-wise potential representing allowed - forbidden layers 
along $z$  and $m=0.067$ is the band edge mass of an electron in GaAs.

In an over-simplified approximation (cf. \cite{14})  the wave function 
just after the barrier can be written as a factorized expression corresponding 
to an approximate separation of variables:
\begin{equation}
\psi_1(y,z)=\frac{C}{\sqrt{k(z)}}
\exp (\frac{i}{\hbar}\int_{0}^zk(z')dz')\psi_0(y).
\label{4a}
\end{equation}
Here $k(z)$ is $z$-momentum and $\psi_0(y)$ is the lowest Landau level 
eigenfunction for motion in the magnetic field
\begin{equation}
\label{3}
\psi_0(y)=(\frac{B\cos \theta}{\pi\hbar})^{1/4}
\exp\left(-\frac{1}{2\hbar }B\cos\theta (y-\tilde{y})^2 \right),
\end{equation}
where $\tilde{y}=<z>\tan \theta$ is
a small shift due to the diagmagnetic term in the perturbation expansion.
$|C|^2=m\Gamma_1$ and $\Gamma_1$ is the imaginary part of the energy of the 
quasi-bound state in the first well computed without the second barrier. 

Using  the Stokes theorem one can show  that the wave 
function inside the QW is expressed through its boundary values as follows
\begin{eqnarray}
\psi(y,z)&&=
\frac{1}{2m}\int 
(G(y,z;y',0)\partial_z\psi_1(y',0)\nonumber\\
&&-\psi_1(y',0)\partial_zG(y,z;y',0))dy',
\label{6d}
\end{eqnarray}
where  $G(y,z;y',z')$ is the Green function for the QW. One can also prove
that this result coincides with the first order correction for wave functions
in the perturbation theory on the tunneling amplitude similar to one used by 
Bardeen in \cite{13}.

Assuming that the QW is sufficiently wide we shall use the standard 
semiclassical approximation according to which the Green
function is represented by a sum over contributions from classical trajectories
($j$) connecting final  $\vec{x}=(y,z)$ and  initial $\vec{x'}=(y',z')$ points
$G(y,z;y',z')=\sum_{j}G_{j}(\vec{x},\vec{x'}),$
and
\begin{equation}
G_{j}(\vec{x},\vec{x'})=\frac{m}{i\sqrt{2\pi\,i k_2 k_1}}
\left |\frac{\partial^2 S_{j}}{\partial t_1 \partial t_2}\right |^{1/2} 
e^{i\frac{S_{j}}{\hbar}-i\frac{\pi}{2}\mu_{j}}.
\label{trajec}
\end{equation}
Here $S_j=S_j(\vec{x},\vec{x'})$ is the classical action
calculated along the classical trajectory $(j)$ connecting initial and final 
points, $\mu_{j}$ is the Maslov index for this trajectory, $k_1$ and $k_2$ 
are the modulus of momentum in the initial and final  points respectively, and 
$t_1$ and $t_2$ are coordinates perpendicular to the trajectory in these 
points. 

In our case,  two modifications have to be made in this standard semiclassical 
formula. (i) In order to to describe the motion inside a well with  
finite width walls it is necessary to multiply the
expression above by the reflection coefficient. (ii) To take into account 
various inelastic processes we shall  multiply 
Eq.~(\ref{trajec}) by a damping factor $\exp (-\Gamma_0 T_j)$
where $T_j$ is the period of the trajectory and $1/\Gamma_0$ plays the role of 
a mean  free time for motion in the QW. In the
following we shall ignore the difference between different kinds of mean free 
paths and in particular the difference between the mean value of the Green 
function and the mean value of the product of the two Green functions which 
is important in describing real experiments (see e.g. \cite{jalab}).

The calculation of the Green function when one of its arguments is 
very close to the boundary of the QW requires special attention. In this
case there are 2 different contributions corresponding to 2 trajectories with  
opposite value of $z$-momentum. The one which hits the boundary  has to be 
multiplied by the reflection coefficient. Therefore
$G(\vec x,\vec{x'} \,)|_{z'=0}=(1+r_1)G_j(\vec x,\vec x_0\,),$
and
$ \partial_{z'} G(\vec x,\vec{x'}\,)|_{z'=0}
=-ip_z(1-r_1)G_j(\vec x,\vec x_0\,),$
where $r_1$ is the reflection coefficient from the left wall 
and $G_j(\vec x,\vec x_0\,)$ is the contribution of the
trajectory starting from the point $\vec x_0=(y',0)$ with $p_z>0$ and ending at 
the point $\vec x$.

The explicit expression for the Green function and the value of the wave 
function in the vicinity of the LH barrier  permits one to make a semiclassical
computation of the wave function at any point inside the QW. We get
\begin{equation}
\psi(y,z)=\sum_{j}\int w_j(y')G_{j}(y,z;y',0)dy'.
\label{psi}
\end{equation}
with
$w_j(y')=iCv_j\sqrt{p_z(j)}\psi_0(y')/2m,$
and
$v_j=(1-r_1)\sqrt{p_z(j)/k_z}+(1+r_1)\sqrt{k_z/p_z(j)}.$
Here $p_z(j)$ is the initial $z$-momentum of trajectory $j$ and 
$k_z=k(0)$ with $k(z)$ from (\ref{4a}).
Under a natural condition $|p(j)-k_z|\ll |p(j)|$ this  expression equals 2 
and we think that this value is a good zeroth order approximation for this 
quantity.

Without the simplifying assumption (\ref{4a}) 
$$C\psi_0(y')/\sqrt{k_z}=\psi_1(y',0) \;\mbox{and}\; 
k_z=\partial_z \log \psi_1(y',0)$$ 
are just the boundary values of 
the initial wave function on the LH barrier to be determined either numerically
or by more refined methods of multidimensional tunneling.

Knowledge of the wave function (\ref{psi}) permits us to compute the
current $\vec {j}_i(\vec{x}_f)=\vec{j}_i(y_f,d)$ at the second interface
of the QW (i.e. at the RH wall):
\begin{equation}
\vec {j}_i(\vec{x}_f)=\frac{1}{2mi} 
(\psi^*(\vec {x}_f){\vec \nabla}\psi(\vec {x}_f)-
\psi(\vec {x}_f){\vec \nabla}\psi^*(\vec {x}_f)).
\end{equation}
When the electron hits the second barrier it has a probability of tunneling 
through the wall. In the same approximation as above the current after the 
barrier, ($\vec{j}_{f}$), differs from the current before it, ($\vec{j}_{i}$),
by a transmission coefficient through this barrier ($t_2$) 
$\vec{j}_{f}=|t_2|^2\vec{j}_{i}.$
The total imaginary part, $\Gamma$, of the energy of a quasi-bound state in the
emitter well equals the total current after the second barrier
\begin{equation}
\Gamma=\int_{S}\,d\vec {\sigma_2}\,\vec {j}_i\,{|t_2|}^2.
\end{equation}
Substituting here the expression (\ref{psi}) for the wave function one gets
an expression for $\Gamma$ as a triple integral over two initial positions 
and one final $y$ coordinate: 
\begin{equation}
\Gamma=\sum_{j,k}
\int dy dy' w_{jk}(y,y')
\int G_{j}(y_f,y)\bar{G}_{k}(y_f,y')dy_f|t_2|^2,
\label{gtotal}
\end{equation}
where 
$w_{jk}(y,y')=(p^{(f)}_z (j)+p^{(f)}_z(k))w_j(y) \bar{w}_k(y')/2m$
and $G_{j}(y_f,y)$ is the contribution (\ref{trajec}) of a trajectory $(j)$ 
which starts at point $(y,0)$ and ends at point $(y_f,d)$.

In the semiclassical approximation $G_j$ is proportional to 
$\exp (iS_j/\hbar)$ and in the formal limit $\hbar\rightarrow 0$
it is natural to perform the integration over all variables by the 
saddle point method. Assuming that the boundary function $\psi_0(y)$ and
other quantities are smooth in the scale of noticeable  changes of the 
Green function, one concludes that in such an approximation the dominant 
contribution to the above integrals will be given by trajectories in the
vicinity of saddle-points trajectories for which the following three conditions 
are fulfilled:
\begin{equation}
\partial S_j(y_f,y)/\partial y_f-
\partial S_k(y_f,y')/\partial y_f=0,
\label{yf}
\end{equation}
\begin{equation}
\partial S_j(y_f,y)/\partial y=0,\;
\partial S_k(y_f,y')/\partial y'=0.
\label{y}
\end{equation}
The first equation means that saddle point trajectories (labeled here by $j$ 
and $k$) should have the same $y$ component of momentum: 
$p^{(f)}_y(j)=p^{(f)}_y(k)$. The equality of the $p_z$ momenta for these 
trajectories then follows from energy conservation. 
But two classical trajectories passing through the same point ($y_f$) and 
having the same momenta at  this point can be either (i) exactly the same 
trajectory or (ii) two different paths on the same  classical trajectory.

The second pair of saddle point equations (\ref{y}) signifies that the
saddle point trajectories should have zero $y$ component of the momentum in 
both points at the LH wall (i.e. they have to be perpendicular 
to the plane $z=0$). The combination of these conditions leads to the important 
conclusion that in the strict semiclassical limit the tunneling current or 
probability of decay is divided into two distinct contributions:
$\Gamma=\Gamma_{(Weyl)}+\Gamma_{(osc)}.$
The first term corresponds to the interference of an arbitrary 
trajectory perpendicular to the plane $z=0$ with itself. 
It has no quick dependence on external 
fields and in real experiments it is 
effectively washed out by taking the second derivative of the current.

The second term, which we denote by $\Gamma_{(osc)}$,  
includes the contributions from different paths on the same self-retracing  
periodic trajectory which connect the LH and RH walls and is perpendicular to 
the plane $z=0$. 
The classical trajectories which give the dominant contribution to 
the resonant tunneling are built from a segment of a periodic orbit which hits
both walls plus an arbitrary number of loops ($n$) around this orbit.

Expanding the actions up to the quadratic terms, expressing the derivatives
of actions through the monodromy matrix, performing the sum over all 
possible repetition numbers one gets
\begin{eqnarray}
&&\Gamma_{(osc)}=\sum_{p}
\Gamma_p  \frac{|t_2^{(tot)}(p)|^2}{1-|R_p|^2}\times
\nonumber\\
&&\sum_{r=1}^{\infty}\frac{R_p^r}{D(r)^{1/2}}
\exp (iS(r)-i\frac{\pi}{2}\mu_p(r)+i\frac{\pi}{4})+c.c.,
\label{final}
\end{eqnarray}
where $D(r)=m_{21}(r)$ and $S(r)=rS_p$.
The sum here is taken over all primitive periodic orbits with $p_y(i)=0$ 
and over up to 2 points of reflection  of this trajectory with the first wall:
$y_0$ and $y'_0$. $S_p$ is the classical action around the periodic orbit 
labeled by $p$ \cite{half},
$m_{ij}(r)$ is the $(i,j)$ monodromy matrix element computed along the 
first barrier for $r$ repetitions of the orbit considered,
$R_p=r_{(tot)}(p)\exp(-\Gamma_0 T_p)$ where $r_{(tot)}(p)$ is
the total reflection coefficient from the walls for one
loop around the given orbit,
$|t_2^{(tot)}(p)|^2=\sum_l|t_2(p,l)|^2\exp(-2\Gamma_0T_0(l))$
is the total tunneling probability through the second barrier weighted by 
damping factors, and $\Gamma_p=(\pi/8)^{1/2}v_p\bar{v}_p'\Gamma_1
\psi_0(y_0)\bar{\psi}_0(y'_0)$.

The formula, Eq.~(\ref{final}), is the main result of the paper. It expresses 
the semiclassical limit of the tunneling current (which is proportional to 
$\Gamma$) as a sum over special  periodic trajectories which (i) connect the 
two walls of the QW and (ii) are perpendicular to the first barrier. 

The important ingredient in deriving the above formula was the assumption 
related to the
applicability of the semiclassical approximation. Whereas for the integration 
over the final coordinate it can easily be argued, the 
requirement that the initial wave function changes slowly in the semiclassical
limit is more difficult to justify. The main problem is that the Landau-type
wave function (\ref{3}) has the term $B\cos \theta y^2/2\hbar$ in the exponent
and both terms, one coming from the initial wave function and one
from the semiclassical Green function, have the same dependance on $\hbar$.
In this case one can proceed as follows. It is easy to check (see (\ref{shift}))
that the requirement of the smoothness of the initial wave function is 
equivalent to the condition 
\begin{equation}
\beta \ll \partial ^2 S(y,y')/\partial y^2=m_{11}/m_{12},
\label{cond}
\end{equation}
where $\beta=B\cos\theta$ and the same condition for the derivative over $y'$.

But the value of the magnetic field in dimensionless units is very small. On the
other hand for generic chaotic systems all elements of the monodromy matrix 
should be of the same order and there is no general reason why the above ratio
should be small. Therefore we expect that for hyperbolic periodic orbits 
with large Lyapunov exponent the condition (\ref{cond}) will be satisfied.
We have checked numerically that the ratio $m_{11}/(m_{12}\beta)$ grows quickly
with increasing $m_{11}$ though for stable and almost stable orbits it 
can be of the order of 1. 

To treat the latter case a slightly different approximation can be used. 
Let us expand the action $S(y,y')$ up to the second order and perform the
integration over $y$ and $y'$ in (\ref{gtotal}) taking into account explicitly 
the dependance of the initial wave function (\ref{3}) on these variables. After
simple algebra we get the same expression as in (\ref{final}) but with 
the pre-factor $D(r)$ and the action $S(r)=rS_p+\Delta S(r)$ substituted by
\begin{eqnarray}
D(r)&=&m_{21}(r)+i\beta(m_{11}(r)+m_{22}(r))-\beta^2m_{12}(r),\nonumber\\
\Delta S(r)&=&\frac{\beta^2}{2D(r)}
(y^2_{0}(m_{22}(r)+i\beta m_{12}(r))\nonumber\\
&+&y^{\prime
2}_{0}(m_{11}(r)+i\beta m_{12}(r))+2y_{0}y'_{0}).
\label{shift}
\end{eqnarray}
These formulae are valid provided the third and higher terms inn the
expansion of the action are small in comparison with quadratic terms. 

When $\beta\rightarrow 0$ this result leads to Eq.~(\ref{final}). But for
stable and almost stable regions where condition (\ref{cond}) is not well 
satisfied Eqs.~(\ref{final}) with  
(\ref{shift}) continue to be a good approximation for $\Gamma$ in 
Eq.~(\ref{gtotal}). In such cases one can (exactly or approximately) split
the Hamiltonian into the sum of two Hamiltonians for longitudinal and transverse
motion and the energy of the former is much bigger than the latter. Under these
conditions (i) the period of motion is defined mainly by the longitudinal motion
and (ii) the transverse motion is close to the motion in a quadratic potential
similar to (\ref{ham}) with $\theta=0$ for a fixed time. But it is known 
\cite{FH} that the semiclassical time-dependent Green function for the motion
in a quadratic potential coincides with the exact Green function and both are 
proportional to the exponential of a function quadratic in its coordinates. 
This means that for stable and almost stable regions there exist good reasons 
why higher than quadratic terms in the expansion of the action on transverse
coordinates should be small and, therefore, Eqs.~(\ref{final})
and (\ref{shift}) are good approximations to the triple integral in 
(\ref{gtotal}). Of course, near bifurcations when by, definition, the quadratic
form in the exponent is degenerate these equations require modification.

There is an interesting limit of the above formulae. It corresponds to 
very clean devices where $\Gamma_0$ is small and the probability of the 
tunneling through the second wall is much bigger than through the fist one. 
In this case $R$ equals the reflection coefficient form the second wall and 
using the relation $|r|^2=1-|t|^2$ one concludes that 
the factor $|t_2^{(tot)}|^2/(1-|R|^2)$ in Eq.~(\ref{final}) tends to 1. One can 
show that all formulae in such a limit coincide with the semiclassical limit of a
simplified model of resonant tunneling discussed in \cite{dom} and \cite{ms} 
which is very convenient for numerical simulation. This model is based on the 
Bardeen transfer matrix \cite{13} according to which the probability of 
tunneling (or the imaginary part of the energy level) is given by
\begin{equation}
\Gamma=2\pi\sum_n |W_n|^2\delta(E-E_n),
\label{bardeen}
\end{equation}
where $E_n$ denotes the exact energy levels in the quantum well and the
coefficients $W_n$ are the matrix elements of the current between the wave
function in the first well ($\psi_1$) and the exact wave function ($\psi_n(y)$) 
in the QW
\begin{equation}
W_n=-\frac{i}{2m}
\int \left (
\frac{\partial \bar \psi_{1}(\vec q\,)}{\partial z}\psi_n(\vec q\,)
-\bar \psi_{1}(\vec q\,)\frac{\partial \psi_n(\vec q\,)}{\partial z}\right )dy,
\label{bard}
\end{equation}
with the $z$-component of point $\vec{q}=(y,z)$ somewhere inside the barrier.

Rewriting Eqs.~(\ref{bardeen}) and (\ref{bard}) through the Green function and 
its $z$-derivatives and using
the same arguments as above it is possible to demonstrate that the 
semiclassical approximation to the resulting expression are given by 
Eqs.~(\ref{final}) with 
$|t_2^{(tot)}|^2/(1-|R|^2)=1$.

The same formulae can be used to derive an approximation similar to 
the Miller modification of the Gutzwiller trace formula \cite{MI}
for a stable orbit. In such a case the monodromy matrix elements for $r$ repetitions 
around a primitive periodic orbit are functions of $\exp (i\omega r)$ where 
$\exp (\pm i\omega)$ are eigenvalues of monodromy matrix. By representing  
Eqs.~(\ref{final}) and (\ref{shift}) as a power series 
$\sum_m \exp (-i\omega r m)C_m$ and performing the summation over $r$ one gets
an expression of $\Gamma$ as a sum over $m$:
\begin{eqnarray}
\Gamma&=&\frac{\pi(\beta \rho)^{1/2}}{1+\beta\rho}\frac{\Gamma_1}{T}
\exp (-\frac{\beta}{2(1+\beta \rho)}(y_0^2+y'^2_0))|v_p|^2\nonumber\\
&\times &\sum_m\frac{1}{2^m m!}\left(\frac{\beta\rho-1}{\beta\rho+1}\right)^m
H_m(\lambda y_{0})H_m(\lambda y'_{0})\nonumber\\
&\times &\sum_N\delta(E-E_{N,m}),
\label{miller}
\end{eqnarray}
where $\rho=m_{12}/\sin \omega$, 
$\lambda = \beta \sqrt{\rho/(\beta^2\rho^2-1)}$
and $E_{N,m}$ are approximate eigenvalues defined as in \cite{MI}:
$S_p(E_{N,m}-\epsilon_{\bot} (m))=2\pi \hbar (N+\mu_p/4)$ and 
$\epsilon_{\bot}=\hbar\omega (m+1/2).$

The index $m$ has the meaning of the number of states in an effective 
transversal Hamiltonian 
$$H(p,y)=\omega(\rho p^2+y^2/\rho)/2.$$
and the resulting expression corresponds to Eqs.~(\ref{bardeen}) and 
(\ref{bard}) in which one approximates $\psi_n(y,z)$ by
a product $\psi(z)\psi_m(y)$ where $\psi_m(y)$ are eigenfunctions of this
Hamiltonian. By truncating this sum at a finite value of $m$ 
one obtains the Miller-type approximation which generalizes the tori 
quantization discussed in \cite{ms}.

In conclusion we have developed a simple semiclassical theory of resonant 
tunneling.  In the strict semiclassical limit the tunneling current is due to 
only special  periodic orbits in the QW  which are perpendicular to the LH 
wall and hit the RH wall of the QW, and their contributions are proportional to 
$(m_{21})^{-1/2}.$ For almost stable orbits one has to use Eq.~(\ref{shift}) 
but the period of oscillations will deviate slightly from that 
of periodic orbits. For big stable regions the Miller-type formula 
(\ref{miller}) is appropriate.

It is a pleasure to acknowledge D. Delande and T. Monteiro for  many useful 
discussions and for presenting us papers \cite{dom} and \cite{ms} prior to 
their publication. The authors 
would also like to thank S.C. Creagh, A. Mouchet, D. Saraga, M. Sieber, 
N.D. Whelan, and D. Ullmo for stimulating discussions.  D.C.R. was supported 
by the Natural Sciences and Engineering Research Council of Canada Fellowship 
(PGS-B award, No. 148961).


\begin{thebibliography}{99}
\bibitem[*]{00} Unit\'e de Recherche des Universit\'es Paris 11 et Paris 6, 
Associ\'ee au CNRS
\bibitem{4} Les Houches Summer School, 1989, {\it Chaos and Quantum Physics}, 
edited by M.J. Giannoni, A. Voros, and J. Zinn-Justin (Elsevier, New York, 
1991)
\bibitem{6} H. Friedrich and D. Wintgen, {\it Phys. Rep.} {\bf 183}, 37 (1989)
\bibitem{8} H.U. Baranger et al, {\it Phys. Rev. Lett.} 
{\bf 70}, 3876 (1993)
\bibitem{jalab} K. Richter et al, {\it Phys. Rep.} {\bf 276}, 1 (1996)
\bibitem{8a} T.M. Fromhold et al, {\it Phys. Rev. Lett.} {\bf 72}, 2608 (1994)
\bibitem{9} T.M. Fromhold et al, {\it Phys. Rev. Lett.} {\bf 75}, 1142 (1995)
\bibitem{10} G. M\"uller et al, {\it Phys. Rev. Lett.} {\bf 75}, 2875 (1995)
\bibitem{9b} P.B. Wilkinson et al, {\it Nature} {\bf 380}, 608 (1996)
\bibitem{11} D.L. Shepelyansky et al, {\it Phys. Rev. Lett.} {\bf 74}, 
2098 (1974)
\bibitem{12} T.S. Monteiro et al, {\it Phys. Rev. E} {\bf 53}, 3369 
(1996)
\bibitem{dom} T.S. Monteiro et al, {\it Phys. Rev.} {\bf B 56}, 3913 (1997)
\bibitem{ms} T.S. Monteiro and D. Saraga, (1997) {\it to be published}
\bibitem{13} J. Bardeen, {\it Phys. Rev. Lett.} {\bf 6}, 57 (1966)
\bibitem{14} M.L. Leadbeater et al, {\it Sem. Sci. Tec.} 
{\bf 6}, 1021 (1991)
\bibitem{half} When $y_0\neq y'_0$ $S_p$ corresponds to a half of periodic 
orbit.
\bibitem{FH} R.P. Feymann and  A.R. Hibbs, {\it Quantum mechanics and Path 
Integrals} (McGraw-Hill Publishing Company, 1965)
\bibitem{MI} W.Miller, {\it J. Chem. Phys.}, {\bf 63}, 996 (1975)
\end{thebibliography}
\end{document}